\documentclass[preprint,showpacs,preprintnumbers,amsmath,amssymb]{revtex4}

\usepackage{graphicx}
\usepackage{dcolumn}
\usepackage{bm}
\usepackage{subfigure,color}

\begin{document}

\title{Light Confinement by a Cylindric Metallic Waveguide in Dense Buffer Gas Environment}

\author{Ulrich Vogl}
\email{vogl@iap.uni-bonn.de}
\author{ Anne Sa\ss}
\author{ Frank Vewinger}
\author{Martin Weitz}%

\affiliation{
Institut f\"ur Angewandte Physik der Universit\"at Bonn, Wegelerstr. 8,
 D-53115 Bonn, Germany}
\author{ Alexander Solovev}
\author{ Yongfeng Mei }
\author{Oliver Schmidt }
\affiliation{Institute for Integrative Nanosciences, IFW Dresden, Helmholtzstr. 20, D-01069 Dresden, Germany }


\date{\today}

\begin{abstract}
We report on the implementation of metallic microtubes in a system of rubidium vapour at 230\,bar of argon buffer gas. The high buffer gas pressure leads to a widely pressure broadened linewidth of several nanometers, interpolating between the sharp atomic physics spectra and the band structure of solid state systems. Tube-like metallic waveguide structures have been inserted in the high pressure buffer gas system, allowing for guiding light in an optical dense gas over a length in the tube of up to 1\,mm. The system holds promise for nonlinear optics experiments and the study of atom-light polariton condensation.
\end{abstract}

\pacs{32.80.-t, 05.30.Jp, 32.70.Jz, 42.50.Fx, 42.82.Et, 78.67.Ch}
\maketitle

Confinement of optical radiation is a key prerequisite in
experiments investigating Bose-Einstein condensation of
polaritons. It allows to tailor the dispersion relation for this
quasi-particles and enhances the obtainable matter-light
interaction. Experimentally, exciton-polariton systems in
microcavity structures gave evidence of quasiparticle condensation
\cite{kasprzak,yamamoto,Giacobino}. In a recent experiment,
thermalization \cite{klaers}, and subsequently Bose-Einstein condensation
\cite{klaers2} of a photon gas in a dye-filled optical microcavity
has been observed, which emphasizes the capacity of this general
approach.

A successful system for light confinement are e.g. hollow core
fibers, which in principle allow strong light-matter coupling over
long distances in a defined volume with a well defined intensity
distribution \cite{Cregan,Gosh,hu,vuletic}. The enhanced
interaction in such systems has proven advantageous for
electromagnetically induced transparency and four-wave-mixing
\cite{russell,Londero} with alkali vapour inside the fiber. Most of these  experiments require careful procedures to coat the inside of the
fiber with certain carbon hydrides to avoid chemical adsorption of
the alkali atoms to the silica bulk of the fiber and is thus not
applicable for hot vapour, but also successful metal-coating has been shown \cite{Fatemi}.

In this paper we investigate metal waveguides in a system of hot
rubidium vapour at 550 Kelvin and 200\,bar of argon buffer
gas pressure. The frequent collisions of the rubidium atoms with
the buffer gas give a pressure broadened linewidth of a few
nanometers, approaching the thermal energy $k_BT$ of the system in
wavelength units. We have shown recently that the frequent
collisions with the buffer gas atoms can lead to thermal
equilibrium of dressed states, i.e. coupled atom-light states
\cite{pra,Chestnov}. The nanosecond lifetimes of alkali excited
states in high pressure buffer gas ($\tau_{nat}\simeq 27 $\,ns for
the case of the 5P state of the rubidium atom) is orders of
magnitude longer than the picoseconds relaxation times of typical
exciton-polariton systems. Thus, there are prospects that atomic
physics based polariton condensation experiments can achieve
longer coherence times than exciton polariton systems. Another
interesting possibility in this system are novel laser cooling
schemes \cite{cooling}, which in a thermodynamic sense may also be
seen as a consequence of coupling internal and external atomic
degrees of freedom in the pressure broadened system. On the other
hand, the high required temperatures of the rubidium-high pressure
buffer gas system (350$^{\circ}$C are required to reach 1\,mbar Rb
vapour pressure, which is the typical equivalent to an optically
dense buffer gas broadened system) make it experimentally
challenging to implement the required optical resonator or
waveguide structures for a particle-like tailoring of the
dispersion relation in the alkali vapour environment. Standard
quartz-based optical fiber structures or mirrors are known to
adsorb alkali vapour at temperatures above 150-200$^{\circ}$C.
Similarly, organic protective coatings become unstable at the used
high temperatures. An interesting alternative possibility is the
use of metallic microstructures, which have higher chemical
stability under the here present conditions. Novel manufacturing
techniques have recently allowed for the fabrication of ultra
thin, rolled metal tubes that can stand the mentioned
preconditions \cite{solovev}. In this work we report on the
implementation of metallic microtubes into a system of rubidium
atoms at 230\,bar argon buffer gas pressure. The ultra thin
microscopic (6\,$\mu$m diameter) metal structures allow to guide
light over a distance of 1\,mm in the high pressure buffer gas
environment. Albeit the harsh environment the waveguide structures
do not show deterioration, even after illuminating with a laser
power of 1 Watt.

A drawback of metallic waveguides is their relatively large loss
compared to dielectric waveguides, which is due to resistive
losses in the metal. Compared to massive metallic structures
thin-walled structures with wall thickness  of order of the
skin-depth can in principle lead to an enhanced transmission
\cite{Takahara,Ghaemi}. A further notable benefit of such metal
structures is that light confinement to diameters below the size
of the wavelength can be achieved, an issue that has allowed for
extraordinarily high transmission through subwavelength diameter
holes \cite{rybczynski}. In an interesting experiment with a
planar configuration with similar subwavelength silver layers the
strong coupling regime could be realized \cite{Hobson}. The silver
layers in this experiment formed a low-Q cavity and the strong
coupling could be realized with the metal boundary condition
providing a stronger confinement than dielectric cavity mirrors.

Let us begin by discussing a few general features of the used
metallic microtube structures. Such microstructures can be
engineered with high precision by depositing, releasing and
rolling up thin metallic membranes \cite{mei1,solovev}. The
fabrication process of these tubes allows for a wide variety of
diameters from nanometer to several micrometer and a length up to
a few millimeter, resulting in aspect ratios of up to 1:10000. The
technique is applicable to a variety of materials including
combinations of semiconductors, metals and oxides
\cite{NatSchmidt}. Figure\,1a shows a Scanning Electron Microscopy
(SEM) image of a rolled up microtube fabricated on silicon
substrates by underetching of photoresist sacrificial layer. The
prepared microtube consists of Ni/Fe/Ag layers and has an internal
diameter of 6$\pm$0.3\,$\mu$m. Figure 1b shows a SEM image of a
microtube transported onto a metallic holder for the integration
into the optical setup. The inset of Figure\,1b indicates
sandwiched Ni/Fe/Ag rolled up layers. A general description for
the production of these waveguides can be found in \cite{mei1},
for the here used samples we applied the following steps: The
Ni/Fe/Ag microtubes were fabricated by electron-beam deposition of
metals onto lithographically patterned photoresist layers. Square
AR-P\,3510 photoresist patterns with the length of 1\,mm were
prepared on 1.5-inch silicon wafer. Photoresist was spin-coated at
3500\,rpm for 35\,s, followed by a soft bake using a hotplate at
90$^{\circ}$C degrees for 1 minute and exposure to UV light with a
Karl Suss MA56 Mask Aligner (410-605\,nm). Patterns were developed
in a 1:1 AR300-35:H2O. Ni/Fe/Ag layers were deposited with
corresponding thicknesses 20/20/20\,nm on the sample tilted to
65$^{\circ}$ from the horizontal line. The tubes were then rolled
up by underetching of the sacrificial photoresist layer in acetone
followed by rinsing in isopropanol. The samples were transferred
into the supercritical point dryer to avoid the tube collapsing
due to the high capillary pressure (surface tension) of the
evaporated solvents. After drying the microtubes were released
using glass micro-capillary and transferred on the metallic
holder. The fabricated structures are then mounted within the high
pressure environment. For the used radiation with wavelength
$\lambda \sim$800\,nm the layer thickness is in the order of the
estimated skin depth of the structure (the skin depth of the inner
silver layer is approximately 12\,nm for the used wavelength and
the thickness of the layer \cite{Wang}).
\begin{figure}[h]
\centerline{
\includegraphics[width=7.5cm]{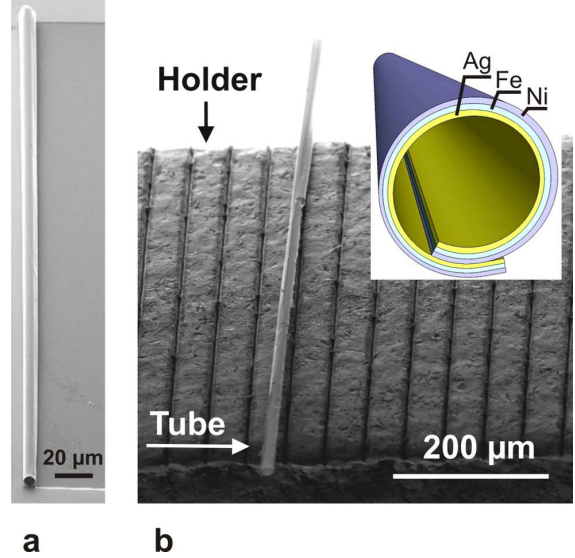}}
 \caption{Single metallic microtube. (a) Scanning electron microscope (SEM) image of  a rolled up metallic microtube on silicon substrate. (b) The microtube  attached to a metallic holder for the integration into the optical setup.
}
\end{figure}

\begin{figure}[t!!!]
\centerline{
\includegraphics[width=7.5cm]{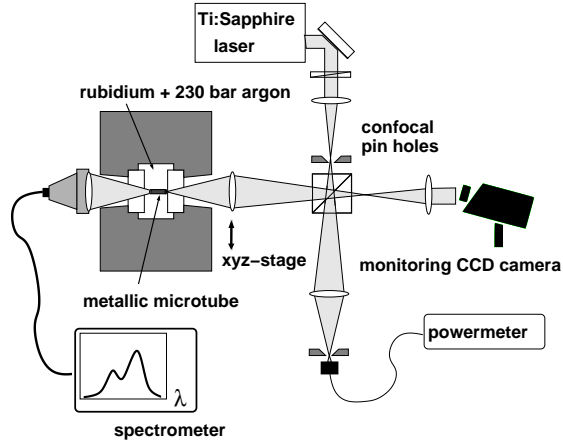}}
 \caption{Experimental setup. With a confocal geometry the position of the laser focus is monitored at a pinhole. A translation stage at the objective allows for lateral scan of the focus position relative to the microtube.
}
\end{figure}
Our experimental apparatus is shown in Figure 2. A steel chamber
with sapphire optical windows is heated to 550\,K, yielding
1\,mbar rubidium vapour pressure (number density
$10^{16}$\,cm$^{-3}$) and 230\,bar argon buffer gas pressure. The
metallic waveguide (Figure 1) is placed on a holder inside the
cell, facing the entrance window of the cell. Tunable laser
radiation from a cw-Titanium-Sapphire laser is directed via a
confocal beam geometry to a movable achromatic lens system and
then focused into the cell. This setup allows to monitor the
location of the laser focus with a control pinhole and
simultaneously steer the microtubes to the focal point of the
objective. The transmitted light was collected and detected
spectrally resolved. In previous experiments \cite{pra} we used a
freely propagating laser beam, which was tightly focused to a beam
waist of $w_0=6\,\mu$m and directed into the vapour cell. The
region of fairly high and uniform intensity is then limited to
twice the Rayleigh length $2z_R=\frac{2\pi
w^2_0}{\lambda}=70\,\mu$m, with the typical used wavelength of
800\,nm. Confinement in a metallic tube over a distance up to
1\,mm thus can enhance the effective interaction length by more
than an order of magnitude.
\begin{figure}
 \centerline{
\includegraphics[width=8.0cm]{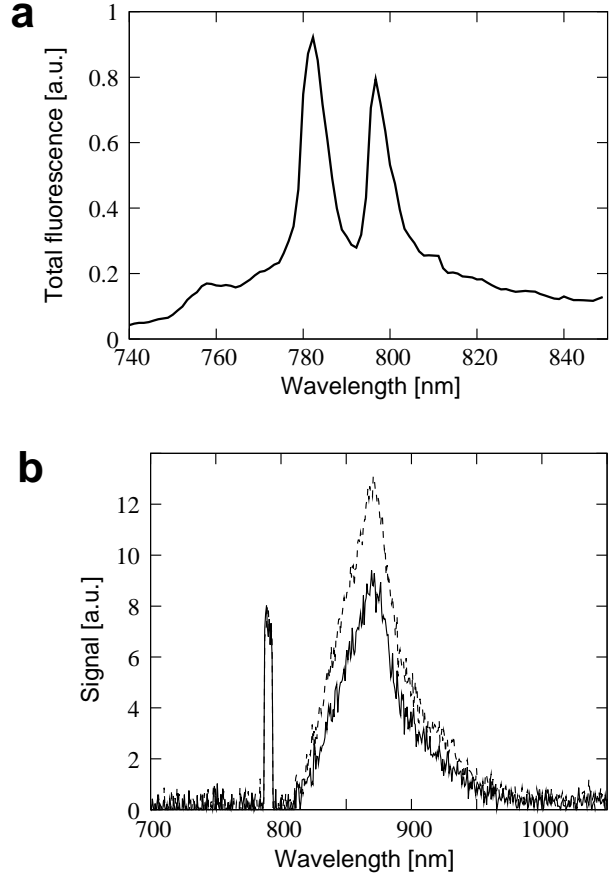}}
    \caption{(a) Spectrum of rubidium with 230\,bar argon buffer gas, recorded in backwards direction. The main features are the two fine structure transitions at 780 and 795\,nm (Rb D2 and D1 lines respectively). The whole spectrum is strongly pressure broadened. (b) Typical spectra of the radiation detected after the cell versus wavelength for an incident wavelength of 785\,nm. One observes a sharp carrier, which is the remainder of the exciting laser, and a broad Stokes-scattered band between 825 and 950\,nm. The solid line shows a spectrum for free transmission. An slightly enhanced Stokes-band is observed when the light is coupled into the tube (dashed line).\label{fig3}}
\end{figure}

In Fig.\,3a we show a typical spectrum in the high pressure buffer
gas environment recorded in backwards direction, i.e. unperturbed
by transmission effects, where the pressure and saturation
broadened rubidium D1 and D2-line resonances are clearly visible.
The observed linewidth (FWHM) is roughly 10\,nm
at the used buffer gas pressure of 230\,bar. The
spectral width here approaches the thermal energy in the heated
gas cell ($k_BT\approx$~20~nm in wavelength units at the D-lines
wavelengths and 550 K temperature) within an order of magnitude.
It is well known that in strongly pressure broadened systems
redistribution of the atomic fluorescence can occur \cite{ct}. In
earlier work we observed that the frequent collisions combined
with an excited state lifetime that exceeds the typical collision
time by three orders of magnitude leads to a complete
redistribution of fluorescence, yielding a spectrum centered
around the two fine structure lines at 780\,nm and 795\,nm in this
system
 \cite{cooling}.

To investigate the possible influence of the microtubes we
analyzed the light in forward direction. A typical corresponding
spectrum of the transmitted light after the cell without metal
tube is shown by the solid line in Fig.\,3b. The used laser
wavelength here was near 785\,nm, i.e. in the vicinity of the
rubidium D-lines, and the visible sharp peak in the spectrum at
the incident wavelength is due to remaining transmitted radiation
at the carrier wavelength. The used optical density is about 3 at resonance, and most of the fluorescence is
multiple reabsorbed and scattered spatially into the full solid
angle and spectrally to the far wings of the fluorescence
spectrum. A main feature of the measured spectra is a relatively
strong and spectrally broad band between 825 and 950\,nm.

Scattered radiation that is closer to resonance within the observed
spectrum is suppressed, as
can be understood by the stronger absorption for near-resonant
light, leading to the observed radiation at the end of the cell
(apart from a remaining part of the incident laser frequency peak)
mainly being far red-shifted radiation. We attribute the observed
red-shifted spectral components in Fig.\,3b to be mainly due to
emission from Rubidium-Argon excimer states \cite{Pascale},
comparable to similar features in Rb-He exciplexes
\cite{Hirano,sell}. Its origin are optical collisions with large
energy exchange, that result in a spectral shift of the
fluorescence photons into the transparent region of the vapour.
As the end of the metal tube is otherwise poorly accessible within
the cell, we use this red shifted spectral band as an indicator
for the influence of the metal tube.

To place the microtubes controlled into the optical path we
initially matched the position of the tube ending and the focus of
the laser beam with the confocal setup shown in Fig.\,2 and
collected the transmitted light. By a lateral scan of the laser
beam in horizontal and vertical direction over the end face of the
metal tube we could clearly identify the tube walls by the
variation in the transmitted light. In Figure\,4 we show the
dependence of the spectral part at the carrier wavelength (dashed
line) and the part in the Stokes-band between 825-950\,nm (solid
line) for a horizontal (panel a) and vertical scan (panel b) over
the waveguide aperture. The shown data were recorded for an
incident wavelength of 775\,nm, where both the transmitted carrier
and the Stokes band have comparable intensity. Clearly visible is
the signature of the tube walls at a distance of circa 6\,$\mu$m
in both the carrier and the Stokes band signal. The observed
attenuation of the total light intensity is small (exept for the
vertical scan, where the beam hits the tube holder an one side).
This alignment and subsequent meaurements were performed with a
laser power up to 1\,W. By monitoring the metal tube no
degradation was observed despite the relatively high light
intensity at the tube entrance of 10$^7$\,W/cm$^2$. This shows
that the immersion of the metal tube in the dense argon gas
provides sufficient heat conduction to prevent melting of the
thinwalled structures. When observing the Stokes band, a slight
enhancement is visible when the light is coupled into the tube, as
can be seen in Fig.~3b. A straightforward explanation for the
variation in the Stokes signal is heating of the waveguide by the
pumping laser, leading to local changes in the density of Argon
and Rubidium. Nevertheless, the overall shape of the spectrum
shows that the waveguide does not disturb the Rubidium-Argon
system strongly, making it thus a promising candidate for
experiments on polariton condensation.

\begin{figure}[t]
\centerline{
\includegraphics[width=7.0cm]{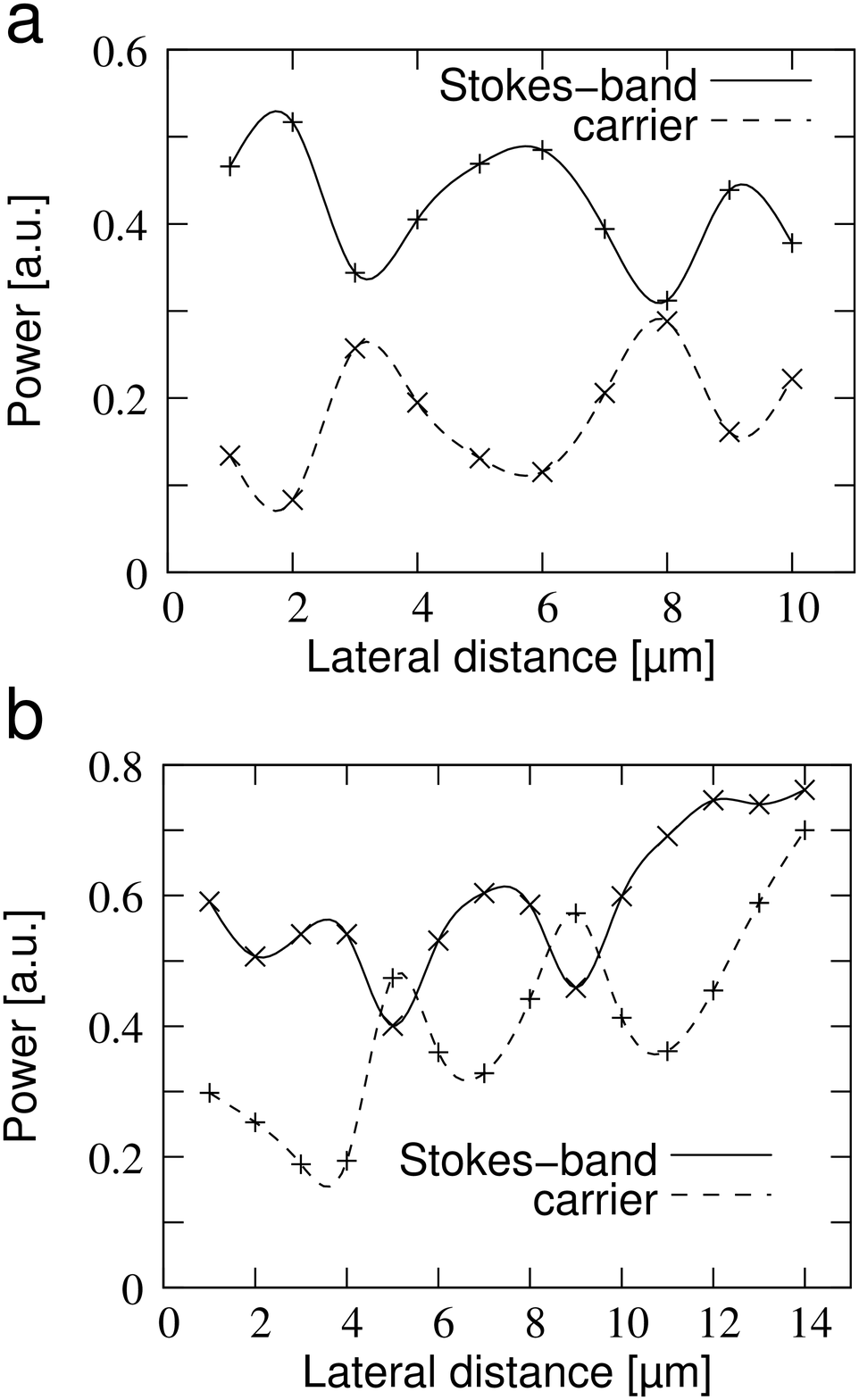}}
\caption{We perfomed a transverse scan of the laser beam over the
tube entrance both in horizontal (a) and vertical (b) direction
(shown are datasets for an incident wavelength of 775\,nm). The
emitted light was collected and spectrally analyzed. Shown are
separately the spectral part of the remaining carrier wavelength
(dashed line) and the Stokes-band between 825-950\,nm for both
cases. We can clearly identify two local drops in the Stokes-band
power at a distance of the tube diameter. The drop of the Stokes
power is accompanied by a peak in the transmitted carrier power.
(The asymmetry in the vertical scan is attributed to the tube
mounting).}
\end{figure}
\begin{figure}[t]
\centerline{
\includegraphics[width=0.45\textwidth]{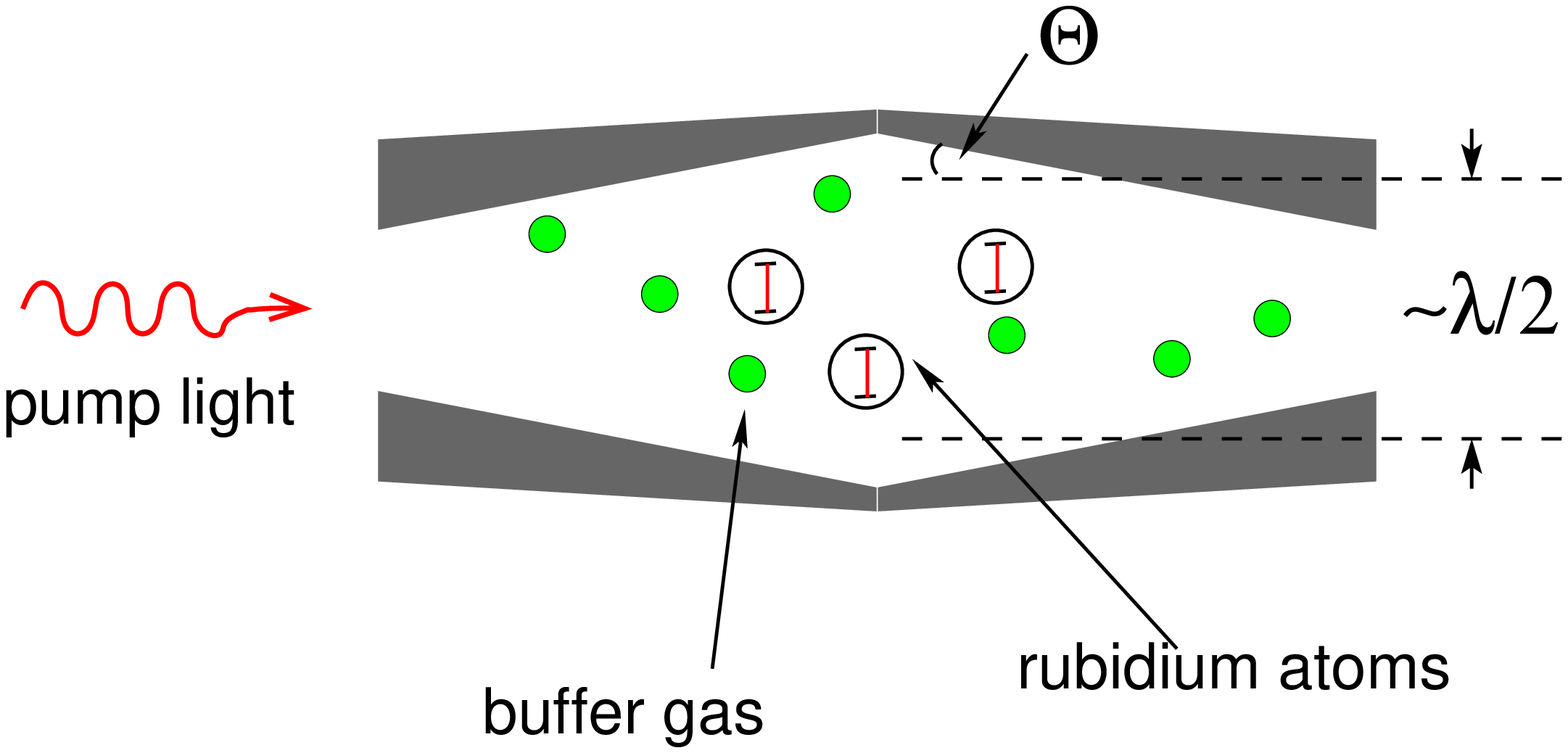}}
\caption{Proposed biconical waveguide design with cut-off within
the redistributed fluorescence spectrum of rubidium. The system
can be optically pumped with short-wavelength light exciting the
pressure broadened rubidium resonances. Subsequent redistribution
of fluorescence towards longer wavelength can accumulate photons
near the cut-off of the waveguide. The diameter of the structure
should be in the order of $\lambda/2$ of the used laser
wavelength. A nonvanishing inner cone angle $\Theta$ of the
biconical structure allows for the tailoring of the allowed modes
inside the cavity \cite{Kravchenko}, and provides a trapping
potential along the weakly confined axis.}
\end{figure}

The here investigated cylindrical metal waveguides could be used to confine ligth to a small volume
 to achieve atom-light polariton
condensation with the waveguide modifying the the dispersion relation of the quasiparticles,
and possibly also acting as a trapping potential.
The combined system of thin-walled metallic
waveguides with rubidium vapour in a high buffer gas environment
exhibits several qualities that make it a promising candidate for
further investigations. The frequent collisions of rubidium atoms
with noble gas atoms under optical radiation can drive coupled
atom-light states towards thermal equilibrium. This can be further
supported by the enhanced interaction within the waveguide. For
the future, the waveguides which have now been designed
cylindrical with relatively large diameter can be tailored to
smaller diameters, which increases the confinement. In particular,
for a diameter of order of $\lambda/2$, the cut-off wavelength
will reach within the redistributed fluorescence spectrum,
similarily as in \cite{klaers}. In this regime we expect that
photons - or, when the strong coupling regime is reached, atom
light polaritons - thermalize above the cavity cut-off, i.e. their
frequencies will be distributed by an amount $k_B/T$ above the
cut-off frequency. The strong transverse confinement makes the
system effectively one-dimensional. Furthermore, by using a
biconical design as indicated in Fig.\,5 a trapping potential can
be realized, that in first order is of the form $V(z)\propto\left|
z \right|$, where $z$ is the direction along the symmetric axis
(in analogy to \cite{klaers}). This linearly confined,
one-dimensional system is expected to support a BEC in the ideal
gas case at sufficiently low temperatures and high densities (the
general requirement is that the trapping potential here should be
more confining than parabolic \cite{Bagnato}). A detection of the
optical radiation could be realized e.g. by collection of the
light leaking out through one of the apertures, or for a tube with
ultrathin walls by light leaking at the waist of the waveguide.

To conclude, we have shown that metallic microtubes can be
integrated in a high pressure buffer gas optical spectroscopy
setup. The material properties allow investigations in
experimental regimes that are not easily accessible with
comparable structures based on silica. Due to the high thermal
conductivity of the metal and the surrounding argon gas, light
intensities of up to 10$^7$\,W/cm$^2$  have been applied without
thermal degradation. These highly versatile structures could pave
the way to new approaches in the strong coupling between light and
matter, or more specific for the investigation of collective
atom-light states. Further investigations will include the study
of smaller structures and waveguides with broken axial symmetry.

Financial support from the DFG within the focused research unit
FOR557 and under the cooperation project 436RUS113/996/0-1 is
acknowledged. The authors acknowledge helpful discussions with
A.~P.~Alodjants and I.~Y.~Chestnov.


\end{document}